\documentclass[12pt]{article}
\usepackage{amsmath}
\usepackage{txfonts,bm,pifont}
\usepackage{cite}
\usepackage{graphicx,xcolor}
\usepackage[dvipdfm,bookmarks=true,bookmarksnumbered=true,colorlinks=true]{hyperref}
\newtheorem{theorem}{Theorem}[section]
\newtheorem{proposition}[theorem]{Proposition}

\newtheorem{lemma}[theorem]{Lemma}
\newtheorem{remark}[theorem]{Remark}
\newtheorem{definition}[theorem]{Definition}
\newcommand{\qed}{\qquad$\square$}
\newcommand{\taus}{\tau^*{}}
\makeatletter
\@addtoreset{equation}{section}
\renewcommand{\theequation}{\@arabic\c@section.\@arabic\c@equation}
\long\def\@makecaption#1#2{
 \vskip 10pt
 \setbox\@tempboxa\hbox{#1. #2}
 \ifdim \wd\@tempboxa >\hsize #1. #2\par \else \hbox
to\hsize{\hfil\box\@tempboxa\hfil}
 \fi}
\makeatother
\begin{document}
\begin{center}
\renewcommand{\baselinestretch}{1.3}\selectfont
\begin{large}
\textbf{Explicit Solutions to the Semi-discrete
 Modified KdV Equation and Motion of Discrete Plane Curves}
\end{large}\\[4mm]
\renewcommand{\baselinestretch}{1}\selectfont
\textrm{\large Jun-ichi Inoguchi$^1$, Kenji Kajiwara$^2$,
Nozomu Matsuura$^3$ and Yasuhiro Ohta$^4$ }\\[2mm]
$^1$: Department of Mathematical Sciences, Yamagata University,\\
 1-4-12 Kojirakawa-machi, Yamagata 990-8560, Japan.\\
 inoguchi@sci.kj.yamagata-u.ac.jp\\
$^2$: Institute of Mathematics for Industry, Kyushu University, \\
744 Motooka, Fukuoka 819-8581, Japan.\\
kaji@imi.kyushu-u.ac.jp
\\
$^3$: Department of Applied Mathematics, Fukuoka University,\\
 Nanakuma, Fukuoka 814-0180, Japan.\\
nozomu@fukuoka-u.ac.jp
\\
 $^4$: Department of Mathematics, Kobe University, Rokko, Kobe 657-8501, Japan.
\\
ohta@math.sci.kobe-u.ac.jp
\\[2mm]
\today
\end{center}
\begin{abstract}
We construct explicit solutions to continuous motion of discrete plane
curves described by a semi-discrete potential modified KdV equation.
Explicit formulas in terms the $\tau$ function are presented.
B\"acklund transformations of the discrete curves are also discussed.
We finally consider the continuous limit of discrete motion of discrete
plane curves described by the discrete potential modified KdV equation
to motion of smooth plane curves characterized by the potential modified
KdV equation.
\end{abstract}



\section{Introduction}
As is well known, many integrable partial differential equations
(integrable systems)
have close relationship to differential geometry. 
In fact, surfaces of specific curvature property 
in 3-dimensional space forms have sine-Gordon type equation 
as the integrability condition of surfaces.
More generally, harmonic maps of conformal 2-manifolds 
into semi-Riemannian symmetric spaces are constructed by 
solutions to 2-dimensional Toda lattice equation (2DTL).

Transformation of solutions to integrable systems 
have origins in classical differential geometry. 
The B{\"a}cklund transformation of the sine-Gordon equation 
are originally formulated as transformations of pseudo-spherical surfaces
in Euclidean $3$-space.

On the other hand, 
substantial progress has been made in the study of 
discretization of integrable systems preserving 
``integrable structure''. 
Motivated by extensive study on discrete integrable systems, 
discretizations of curves and surfaces have been recently 
studied actively. 

This paper concerns with geometry of discrete curves 
in terms of semi-discrete integrable systems.
In \cite{DS1,DS2,DS3}, Doliwa and Santini
introduced continuous motion of discrete 
curves in 3-sphere described by 
the Ablowitz-Ladik hierarchy \cite{Ablowitz-Ladik:1975}.
The semi-discrete potential mKdV equation was deduced as the simplest case.
Hoffmann and Kutz \cite{HK} introduced the notion of 
discrete curvature for plane discrete curves.
Using the discrete curvature, they 
deduced the semi-discrete mKdV equation from 
continuous motion of plane discrete curves.

In our previous works \cite{IKMO:discrete_curve, Matsuura}, we
have studied discrete motions of plane discrete curves in purely 
Euclidean geometric manner. The compatibility condition of a
discrete motion is the discrete potential mKdV equation proposed by 
Hirota \cite{Hirota:dpmKdV}. In discrete differential geometric setting,
the primal geometric object is the potential function rather than curvature
(see \cite{Matsuura}).
Note that potential function coincides with the turning angle function in smooth 
curve theory. We have constructed explicit solutions of discrete motions of 
plane discrete curves in \cite{IKMO:discrete_curve}.

As a continuation of the previous works, in this paper we study 
continuous motions of plane discrete curves in terms of potential function.
The purpose of the present paper is to construct 
explicit solutions to continuous motions of 
plane discrete curves by using the so-called $\tau$ function. 
Moreover we shall give B\"acklund transformations of continuous 
motions of plane discrete curves. 
The discrete curvature functions and 
the semi-discrete mKdV equation discussed in \cite{HK} 
are recovered from our results. 

We have been working on three categories of curves motions:
(1) continuous motions of plane smooth curves,
(2) continuous motions of plane discrete curves and
(3) discrete motions of plane discrete curves.
In this paper we investigate the relationship of these three motions,
and show that these motions are connected by appropriate continuous
limiting procedure.

This paper is organized as follows.  After recalling the requisite facts
on the geometry of plane discrete curves and their continuous motion in
Section 2, we prepare a representaion formula for continuous motion of
plane discrete curves in terms of $\tau$ function.  This representation
enable us to give explicit parametrization of motions determined by
multi-solitons as well as multi-breathers in the next Section 4.

As we have mentioned before, B{\"a}cklund transformation is a
fundamental and effective tool for construction of solutions. In Section
5, we extend B{\"a}cklund transformations of plane discrete curves
studied in our previous work \cite{IKMO:discrete_curve} to those of
continuous motions.  In particular, we give a new formula for
B{\"a}cklund transformations on the semi-discrete potential mKdV equation.

In the final section, we shall discuss continuous limits of motions of
plane discrete curves.  More precisely, first we shall investigate
continuous limits of discrete motions of plane discrete curves to
continuous motion of those. Next we study continuous limits of
continuous motions of plane discrete curves to continuous motions of
plane smooth curves.  It should be emphasized that these limiting
procedure preserve solutions of equations.  More precisely, we shall
show that these limiting procedure preserve soliton type solutions. This
is confirmed by careful analysis of $\tau$ functions.  Appendix will be
devoted to detailed computations of bilinear equations for our use.
 
In a separate publication \cite{FIKMO}, we study discrete hodograph
transformations and apply those to obtain discretizations of some
integrable systems associated with continuous motions of plane smooth
curves.

\section{Continuous Motion of Plane Discrete Curves}
We start with the following definition.
\begin{definition}\rm
 A map $\gamma:~\mathbb{Z}\rightarrow \mathbb{R}^2;~l\mapsto \gamma_l$ is said to be a
 \textit{discrete curve} of constant segment length $\epsilon$ if
\begin{equation}
 \left|\frac{\gamma_{l+1} - \gamma_l}{\epsilon}\right| = 1.
\label{iso0}
\end{equation}
\end{definition}
\par\bigskip
\noindent We introduce the \textit{angle function} $\psi_l$ of a discrete curve $\gamma$ by
\begin{equation}
 \frac{\gamma_{l+1} - \gamma_l}{\epsilon}=\left[\begin{array}{l}\cos\psi_l \\\sin\psi_l \end{array}\right].\label{discrete_angle}
\end{equation}
A discrete curve $\gamma$ satisfies
\begin{equation}
 \frac{\gamma_{l+1} - \gamma_{l}}{\epsilon} = R(K_l)~\frac{\gamma_{l} - \gamma_{l-1}}{\epsilon},
\label{discrete_FS}
\end{equation}
for $K_l=\psi_{l}-\psi_{l-1}$, where $R(K_l)$ denotes the rotation matrix given by
\begin{equation}
 R(K_l)=\left[\begin{array}{cc}\cos K_l & -\sin K_l\\\sin K_l & \cos K_l\end{array}\right].
\end{equation}
We consider the following motion of discrete curves:
\begin{align}
 \frac{d\gamma_l}{ds}=\frac{1}{\cos\frac{K_l}{2}}~R\left(-\frac{K_l}{2}\right)~\frac{\gamma_{l+1}-\gamma_{l}}{\epsilon}.\label{gamma_s}
\end{align}
Then from the compatibility condition of (\ref{discrete_FS}) and
(\ref{gamma_s}), there exists a potential function $\theta_l$ such that
\begin{equation}
 \psi_l = \frac{\theta_{l+1}+\theta_l}{2},\quad K_l =  \frac{\theta_{l+1}-\theta_{l-1}}{2},\label{Psi_and_theta}
\end{equation}
and it follows that from the isoperimetric condition (\ref{iso0}) that $\theta_l$ satisfies
\begin{equation}
\frac{d\theta_l}{ds} = \frac{2}{\epsilon}\tan\left(\frac{\theta_{l+1}-\theta_{l-1}}{4}\right). \label{sdpmKdV}
\end{equation}
Equation (\ref{sdpmKdV}) is called the 
\textit{semi-discrete potential modified}
\textit{KdV} (\textit{mKdV}) \textit{equation}.

Hoffmann and Kutz \cite{Hoffmann:LN,HK} introduced (\ref{gamma_s}) as the edge tangential flow of
discrete plane curves, which was deduced by discretizing the curvature function of motion of plane
smooth curves. We note that Doliwa and Santini formulated in \cite{DS1,DS2,DS3} the integrable
motion of discrete curves in 3-sphere described by the Ablowitz-Ladik hierarchy, where the
semi-discrete potential mKdV equation (\ref{sdpmKdV}) arises as the simplest case.
Their formulation includes the motion of plane curves as a limiting case.
%

\section{Representation Formula in terms of $\tau$ Function}\label{sec:tau}
In this section, we present a representation formula for curve motions in terms of $\tau$ function.
We also give explicit $\tau$ functions which correspond to soliton and breather solutions.

Let $\tau_l=\tau_l(s;y)$ be a complex function dependent on the discrete variable $l$ and two
continuous variables $s$ and $y$, satisfying the following system of bilinear equations:
\begin{align}
& D_s~\tau_l\cdot\tau^*_l=\frac{1}{2\epsilon}\left(\tau^*_{l-1}\tau_{l+1}-\tau_{l+1}^*\tau_{l-1}\right),\label{bl1}\\[2mm]
& \tau_l\tau^*_l=\frac{1}{2}\left(\tau^*_{l-1}\tau_{l+1}+\tau^*_{l+1}\tau_{l-1}\right),\label{bl2}\\[2mm]
& \frac{1}{2}D_sD_y~\tau_l\cdot\tau_l = -\tau^*_{l+1}\tau^*_{l-1},\label{bl3}\\[2mm]
& D_y~\tau_{l+1}\cdot\tau_l = -\epsilon\tau^*_{l+1}\tau^*_l.\label{bl4}
\end{align}
Here, ${}^*$ denotes the complex conjugate,  $D_s$, $D_y$ are the Hirota's \textit{bilinear differential
operators} ($D$-operators) defined by
\begin{equation}
 D_s^iD_y^j~f\cdot g = \left.(\partial_s-\partial_{s'})^i(\partial_y-\partial_{y'})^j~f(s,y)g(s',y')\right|_{s=s',y=y'}~.
\end{equation}
We refer to \cite{Hirota:book} for calculus of $D$-operators. The functions satisfying the bilinear
equations are called the \textit{$\tau$ functions}. 
%
\begin{theorem}\label{thm:tau_formula}
 Let $\tau_l$ be a solution to eqs.(\ref{bl1})--(\ref{bl4}). Define a real function
$\theta_l(s;y)$ and an $\mathbb{R}^2$-valued function $\gamma_l(s;y)$ by
\begin{align}
& \theta_l(s;y):=\frac{2}{\sqrt{-1}}\log\frac{\tau_l}{\tau_l^*},\\
&\gamma_l(s;y):=\left[\begin{array}{c} 
{\displaystyle -\frac{1}{2}\left(\log \tau_l\tau_l^*\right)_y}\\[2mm]
{\displaystyle \frac{1}{2\sqrt{-1}}\left(\log \frac{\tau_l}{\tau_l^*}\right)_y}
\end{array}\right].
\end{align}
Then for any $s,y\in\mathbb{R}$ and $l\in\mathbb{Z}$, the functions $\theta_l=\theta_l(s;y)$ and
$\gamma_l=\gamma_l(s;y)$ satisfy (\ref{iso0}), (\ref{discrete_FS}) (\ref{gamma_s}) and (\ref{sdpmKdV}).
\end{theorem}
\noindent{\bf Proof.} Express $\gamma_l={}^t(X_l,Y_l)$. From (\ref{bl4}) and its complex conjugate we have
\begin{equation}
 \left(\log\frac{\tau_{l+1}}{\tau_l}\right)_y = -\epsilon~\frac{\tau_{l+1}^*\tau_l^*}{\tau_{l+1}\tau_l},\qquad
\left(\log\frac{\tau_{l+1}^*}{\tau_l^*}\right)_y  = -\epsilon~\frac{\tau_{l+1}\tau_l}{\tau_{l+1}^*\tau_l^*}.\label{th1:proof1}
\end{equation}
Adding these two equations we obtain
\begin{equation}
 \left(\log\tau_{l+1}\tau_{l+1}^*\right)_y -  \left(\log\tau_{l}\tau_{l}^*\right)_y 
=-\epsilon\left(\frac{\tau_{l+1}^*\tau_l^*}{\tau_{l+1}\tau_l}
+ \frac{\tau_{l+1}\tau_l}{\tau_{l+1}^*\tau_l^*}\right),
\end{equation}
which yields
\begin{equation}
\frac{X_{l+1} - X_l}{\epsilon} = 
\cos\psi_l,\quad \psi_l
= \frac{1}{\sqrt{-1}}
\log\left(\frac{\tau_{l+1}\tau_l}{\tau_{l+1}^*\tau_l^*}\right)
= \frac{\theta_{l+1}  + \theta_{l} }{2}.
\end{equation}
Subtracting the second equation from the first equation in eq.(\ref{th1:proof1}) we have
\begin{eqnarray*}
\frac{Y_{l+1} - Y_l}{\epsilon} = \sin\psi_l.
\end{eqnarray*}
Therefore we obtain
\begin{equation}
\frac{\gamma_{l+1} -\gamma_l}{\epsilon}
= \left[\begin{array}{c}\smallskip\cos\psi_l \\\sin\psi_l \end{array}\right].\label{segment_l}
\end{equation}
which gives eq.(\ref{iso0}). Next, from eq.(\ref{segment_l}) we see that 
\begin{equation}
\frac{\gamma_{l+1} -\gamma_l}{\epsilon}
=R(\psi_l - \psi_{l-1})~\frac{\gamma_{l} -\gamma_{l-1}}{\epsilon},\quad
\psi_l - \psi_{l-1} =\frac{\theta_{l+1}-\theta_{l-1}}{2}= K_l,
\end{equation}
which is nothing but eq.(\ref{discrete_FS}). In order to show (\ref{gamma_s}), we identify
$\mathbb{R}^2$ as $\mathbb{C}$. Then by using (\ref{discrete_angle}) and (\ref{Psi_and_theta}),
we see that (\ref{gamma_s}) is rewritten as
\begin{equation}
  \cos\frac{K_l}{2}~\dot\gamma_l=e^{-\sqrt{-1}\frac{K_l}{2}}~\frac{\gamma_{l+1}-\gamma_l}{\epsilon}
=e^{\sqrt{-1}\frac{\theta_{l+1}+2\theta_l+\theta_{l-1}}{4}}.\label{prop:eq1}
\end{equation}
We have
\begin{align}
\cos\frac{K_l}{2}
&=\frac{1}{2}\left[e^{\sqrt{-1}\frac{\theta_{l+1}-\theta_{l-1}}{4}}+e^{-\sqrt{-1}\frac{\theta_{l+1}-\theta_{l-1}}{4}}\right]
=\frac{1}{2}\left[\left(\frac{\tau_{l+1}\tau^*_{l-1}}{\tau^*_{l+1}\tau_{l-1}}\right)^{1/2}
+\left(\frac{\tau^*_{l+1}\tau_{l-1}}{\tau_{l+1}\tau^*_{l-1}}\right)^{1/2}\right]\notag\\
&=\frac{1}{2}~\frac{\tau_{l+1}\tau^*_{l-1} + \tau_{l-1}\tau^*_{l+1}}{\left|\tau_{l+1}\tau_{l-1}\right|}
=\frac{\tau_{l}\tau^*_{l} }{\left|\tau_{l+1}\tau_{l-1}\right|},
\end{align}
where we have used (\ref{bl2}). Noticing that
\begin{equation}
\gamma_l= X_l+\sqrt{-1}Y_l
=-\left(\log\tau^*_l\right)_y,
\end{equation}
the left hand side of (\ref{prop:eq1}) can be rewritten by using (\ref{bl3}) as
\begin{align*}
 &\cos\frac{K_l}{2}~\frac{d\gamma_l}{ds}
=\frac{\tau_{l}\tau^*_{l} }{\left|\tau_{l+1}\tau_{l-1}\right|}\times(-1)\left(\log\tau^*_l\right)_{ys}
=-\frac{\tau_{l}\tau^*_{l} }{\left|\tau_{l+1}\tau_{l-1}\right|}~\frac{\frac{1}{2}D_sD_y~\tau^*_l\cdot\tau^*_l}{\left(\tau^*_l\right)^2}
=\frac{\tau_{l+1}\tau_{l-1}}{\left|\tau_{l+1}\tau_{l-1}\right|}~\frac{\tau_l}{\tau^*_l}
\\
&=e^{\sqrt{-1}\frac{\theta_{l+1}+2\theta_l+\theta_{l-1}}{4}},
\end{align*}
which implies (\ref{prop:eq1}). Finally, the semi-discrete potential mKdV equation (\ref{sdpmKdV}) can be
derived by dividing (\ref{bl1}) by (\ref{bl2}). \qed
\section{Explicit Solutions}
We now present explicit formulas for the $\tau$ function which correspond to multi-soliton and
multi-breather solutions to the bilinear equations, respectively.

\begin{theorem}\label{thm:solutions}
For $N\in\mathbb{Z}_{\geq 0}$, consider the $\tau$ function
 \begin{equation}
 \tau_l(s;y) = \exp\left[-\left(s+\epsilon l\right)y\right]~\det\left(f_{j-1}^{(i)}\right)_{i,j=1,\ldots,N},\label{tau}
\end{equation}
\begin{equation}
 f_n^{(i)} = \alpha_i p_i^n (1-\epsilon p_i)^{-l} e^{\frac{p_i}{1-\epsilon^2p_i^2}s + \frac{1}{p_i}y}
+\beta_i (-p_i)^n (1+\epsilon p_i)^{-l} e^{-\frac{p_i}{1-\epsilon^2p_i^2}s - \frac{1}{p_i}y},\label{tau:entries}
\end{equation}
where $\alpha_i$, $\beta_i$ and $p_i$ ($i=1,\ldots,N$) are parameters.
\begin{enumerate}
 \item Choosing the parameters as
\begin{equation}\label{param:soliton}
p_i,\ \alpha_i\in\mathbb{R},\quad \beta_i\in\sqrt{-1}\mathbb{R}\quad (i=1,\ldots,N),
\end{equation}
then $\tau_l$ satisfies the bilinear equations (\ref{bl1})--(\ref{bl4}). This gives the
$N$-soliton solution to (\ref{sdpmKdV}).
 \item Taking $N=2M$, and choosing the parameters as
\begin{equation}\label{param:breather}
\begin{array}{l}\medskip
 {\displaystyle  p_i,\ \alpha_i,\ \beta_i \in\mathbb{C}\quad (i=1,\ldots,2M),
\quad p_{2r}=p_{2r-1}^*\quad (r=1,\ldots,M),}\\
{\displaystyle \alpha_{2r}=\alpha_{2r-1}^*,\quad  \beta_{2r}=-\beta_{2r-1}^*\quad (r=1,\ldots,M),}
\end{array}
\end{equation}
then $\tau_l$ satisfies the bilinear equations (\ref{bl1})--(\ref{bl4}). This gives the
$M$-breather solution to (\ref{sdpmKdV}).
\end{enumerate}
\end{theorem}

In order to prove Theorem \ref{thm:solutions}, we first consider the following ``generic'' $\tau$ function
and system of bilinear equations. Then Theorem \ref{thm:solutions} is derived by applying the
reduction.
\begin{proposition}\label{prop:generic}
Let $\sigma^{k}_{l,m}=\sigma^{k}_{l,m}(u,v;y)$ is a function depending on three discrete independent
 variables $k,l,m\in\mathbb{Z}$ and three continuous independent variables $u,v,y\in\mathbb{R}$
 defined by
\begin{equation}
 \sigma^k_{l,m}(u,v;y) = \det\left(f_{k+j-1}^{(i)}(l,m)\right)_{i,j=1,\ldots,N},\label{sigma}
\end{equation}
\begin{equation}
 f_k^{(i)}(l,m) = \alpha_i p_i^k (1-ap_i)^{-l}(1-bp_i)^{-m} e^{\frac{u}{1-ap_i}+\frac{v}{1-bp_i} + \frac{1}{p_i}y}
+\beta_i q_i^k (1-aq_i)^{-l}(1-bq_i)^{-m} e^{\frac{u}{1-aq_i}+\frac{v}{1-bq_i} + \frac{1}{q_i}y}\label{sigma:entries}
\end{equation}
where $a$, $b$, $\alpha_i$, $\beta_i$, $p_i$ and $q_i$ ($i=1,\ldots,N$) are parameters. Then $\sigma_{l,m}^k$
 satisfies the following bilinear equations:
\begin{align}
 & \left(D_u-1\right)~\sigma^{k-1}_{l,m}\cdot\sigma^{k}_{l,m}=-\sigma^{k}_{l+1,m}\sigma^{k-1}_{l-1,m},\label{bl5}\\[2mm]
 & \left(D_v-1\right)~\sigma^{k-1}_{l,m}\cdot\sigma^{k}_{l,m}=-\sigma^{k}_{l,m+1}\sigma^{k-1}_{l,m-1},\label{bl6}\\[2mm]
 & b\sigma^{k+1}_{l,m+1}\sigma^{k}_{l+1,m} - a\sigma^{k+1}_{l+1,m}\sigma^{k}_{l,m+1} + (a-b)\sigma^{k+1}_{l+1,m+1}\sigma^{k}_{l,m}=0,\label{bl7}\\[2mm]
 & \frac{1}{2}D_uD_y~\sigma^{k}_{l,m}\cdot\sigma^{k}_{l,m} = a(\sigma^{k}_{l,m})^2 - a\sigma^{k+1}_{l+1,m}\sigma^{k-1}_{l-1,m},\label{bl8}\\[2mm]
 & \frac{1}{2}D_vD_y~\sigma^{k}_{l,m}\cdot\sigma^{k}_{l,m} = b(\sigma^{k}_{l,m})^2 - b\sigma^{k+1}_{l,m+1}\sigma^{k-1}_{l,m-1},\label{bl9}\\[2mm]
 & \left(D_y-a\right)~\sigma^{k}_{l+1,m}\cdot \sigma^{k}_{l,m} = -a \sigma^{k+1}_{l+1,m}\sigma^{k-1}_{l,m}.\label{bl10}
\end{align}
\end{proposition}
\noindent{\bf Proof of Theorem \ref{thm:solutions}}\quad 
We show that Theorem \ref{thm:solutions} holds from Proposition \ref{prop:generic}. We impose the
reduction conditions on $\sigma^k_{l,m}$ as
\begin{align}
 &\sigma^k_{l+1,m+1}=B\sigma^k_{l,m},\label{reduction_condition1}\\
 &\sigma^{k+1}_{l,m}=C\sigma^*{}^k_{l,m},\quad C\in\mathbb{R}\label{reduction_condition2}
\end{align}
where $B,C$ are constants. Then putting $b=-a$, the bilinear equations (\ref{bl5})--(\ref{bl10}) are reduced to
\begin{align}
 & \left(D_u-1\right)~\sigma^*_l\cdot\sigma_l=-\sigma_{l+1}\sigma^*_{l-1},\label{bl11}\\[2mm]
 & \left(D_v-1\right)~\sigma^*_{l}\cdot\sigma_l=-\sigma_{l-1}\sigma^*_{l+1},\label{bl12}\\[2mm]
 & \sigma^*_{l-1}\sigma_{l+1} + \sigma^*_{l+1}\sigma_{l-1} - 2\sigma^*_{l}\sigma_{l}=0,\label{bl13}\\[2mm]
 & \frac{1}{2}D_uD_y~\sigma_{l}\cdot\sigma_{l} = a(\sigma_{l})^2 - a\sigma^*_{l+1}\sigma^*_{l-1},\label{bl14}\\[2mm]
 & \frac{1}{2}D_vD_y~\sigma_{l}\cdot\sigma_{l} = -a(\sigma_{l})^2 + a\sigma^*_{l-1}\sigma^*_{l+1},\label{bl15}\\[2mm]
 & \left(D_y-a\right)~\sigma_{l+1}\cdot \sigma_{l} = -a \sigma^*_{l+1}\sigma^*_{l},\label{bl16}
\end{align}
respectively. Here we have used (\ref{reduction_condition1}) and (\ref{reduction_condition2}) to
eliminate the $m$- and $k$-dependence, respectively, and denoted $\sigma^k_{l,m}=\sigma_l$. We next
consider the specialization of continuous independent variables
\begin{equation}
 u = cs,\quad v=-cs,\quad c\in\mathbb{R}.\label{specialization1}
\end{equation}
Then, subtracting (\ref{bl12}) from (\ref{bl11}) we have
\begin{equation}
 D_s~\sigma^*_l\cdot\sigma_l = c\left(\sigma_{l-1}\sigma^*_{l+1} - \sigma_{l+1}\sigma^*_{l-1} \right).\label{bl17}
\end{equation}
Similarly, we get from (\ref{bl14}) and (\ref{bl15}) 
\begin{equation}
 D_sD_y~\sigma_{l}\cdot\sigma_{l} = 4ac\left\{(\sigma_{l})^2 - \sigma^*_{l-1}\sigma^*_{l+1}\right\}.\label{bl18}
\end{equation}
Putting
\begin{equation}
  a=\epsilon,\quad c = \frac{1}{2\epsilon}, \label{specialization2}
\end{equation}
and introducing $\tau_l$ by 
\begin{equation}
 \tau_l = e^{-(s+\epsilon l)y}\sigma_l,
\end{equation}
the bilinear equations (\ref{bl17}), (\ref{bl13}), (\ref{bl18}), (\ref{bl16}) are reduced to
(\ref{bl1}), (\ref{bl2}), (\ref{bl3}), (\ref{bl4}), respectively.  Let us next realize the reduction
conditions (\ref{reduction_condition1}) and (\ref{reduction_condition2}) by imposing suitable
restriction on parameters of solution. We put
\begin{equation}
 q_i = -p_i\quad (i=1,\ldots,N),\quad b=-a.
\end{equation}
Then it is easy to verify that the entries of the determinant satisfy
\begin{equation}
 f_k^{(i)}(l+1,m+1)=\frac{1}{1-a^2p_i^2}~f_k^{(i)}(l,m),
\end{equation}
so that the condition (\ref{reduction_condition1}) is realized as
\begin{equation}
 \sigma^k_{l+1,m+1} = \prod_{i=1}^N\frac{1}{1-a^2p_i^2}~\sigma^k_{l,m}.
\end{equation}
As for the condition (\ref{reduction_condition2}), we have to consider the cases (1) and (2) in
Theorem \ref{thm:solutions} separately:

\noindent Case (1). We impose the condition (\ref{param:soliton}). Then we see that
\begin{equation}
 f_{k+1}^{(i)}(l,m)=p_i~f_k^{(i)*}(l,m),
\end{equation}
and so
\begin{equation}
 \sigma^{k+1}_{l,m}=C~\sigma^k_{l,m},\quad C=\prod_{i=1}^N p_i\in\mathbb{R}.
\end{equation}
\noindent Case (2). We impose the condition (\ref{param:breather}). Then we see that
\begin{equation}
 f_{k+1}^{(2r)}(l,m)=p_{2r-1}^*f_{k}^{(2r-1)*}(l,m),\quad
 f_{k+1}^{(2r-1)}(l,m)=p_{2r}^*f_{k}^{(2r)*}(l,m),
\end{equation}
and so
\begin{equation}
  \sigma^{k+1}_{l,m}=C~\sigma^*{}^k_{l,m},\quad C=(-1)^M\prod_{r=1}^M |p_{2r}|^2\in\mathbb{R}.
\end{equation}

Finally, putting $m=0$ without loss of generality and applying the specialization (\ref{specialization1})
and (\ref{specialization2}), (\ref{sigma:entries}) is rewritten as
\begin{align*}
 & f_k^{(i)}(l,0) = \alpha_i p_i^k(1-\epsilon p_i)^{-l}e^{\frac{s}{2\epsilon}\left(\frac{1}{1-\epsilon p_i}-\frac{1}{1+\epsilon p_i}\right) + \frac{1}{p_i}y}
+\beta_i (-p_i)^k(1+\epsilon p_i)^{-l}e^{\frac{s}{2\epsilon}\left(\frac{1}{1+\epsilon p_i}-\frac{1}{1-\epsilon p_i}\right) - \frac{1}{p_i}y}\\
&= \alpha_i p_i^k(1-\epsilon p_i)^{-l}e^{\frac{p_i}{1-\epsilon^2 p_i^2} s+ \frac{1}{p_i}y}
+\beta_i (-p_i)^k (1+\epsilon p_i)^{-l}e^{-\frac{p_i}{1-\epsilon^2 p_i^2} s - \frac{1}{p_i}y},
\end{align*}
which is equivalent to (\ref{tau:entries}). Therefore we have derived Theorem \ref{thm:solutions}
from Proposition \ref{prop:generic}.\qed

\par\bigskip

The bilinear equations in Proposition \ref{prop:generic} are reduced to the quadratic identities of
determinants (Pl\"ucker relations). In particular, (\ref{bl7}) and (\ref{bl10}) have already
appeared in \cite{IKMO:discrete_curve}. Moreover, by the symmetry between the set of variables
$(l,u)$ and $(m,v)$ in $\sigma^k_{l,m}$, it suffices to show only (\ref{bl5}) and (\ref{bl8}). These
bilinear equations will be proved in the Appendix.

\begin{remark}\rm
In the $\tau$ function in Theorem \ref{thm:solutions}, the parameter dependence of the time
evolution in entries of the Casorati determinant have singularities different from $0$ and $\infty$.
These types of singularities can be seen in the solutions of equation of principal chiral fields, \textit{i.e.},
harmonic maps of conformal $2$-manifolds into compact Lie groups \cite{Jimbo-Miwa, Uhlenbeck,
Zakharov-Mikhailov} and Maxwell-Bloch equation \cite{Kakei-Satsuma:MB}.
\end{remark}

\begin{remark}\rm
By introducing $u_l:=\frac{\epsilon}{2}\frac{d\theta_l}{ds}$, the semi-discrete potential mKdV equation
(\ref{sdpmKdV}) can be transformed to the semi-discrete mKdV equation
\begin{equation}
 \frac{du_l}{ds'} = \left(1+u_l^2\right)(u_{l+1}-u_{l-1}),\label{eqn:sdmKdV}
\end{equation}
where we put $s=2\epsilon s'$ for convenience. An auxiliary linear problem for (\ref{eqn:sdmKdV})
 is given by \cite{DS1}
\begin{equation}
\Phi_{l+1}=\frac{1}{\sqrt{1+u_l^2}}
\left[\begin{array}{cc} \lambda & \lambda^{-1}u_l\\-\lambda u_l & \lambda^{-1}
      \end{array}\right]\Phi_l,\quad
\frac{d}{ds'}\Phi_l 
= \left[\begin{array}{cc} \frac{\lambda^2-\lambda^{-2}}{2} &    u_l+\lambda^{-2}u_{l-1}\\
-u_l-\lambda^2 u_{l-1} & -\frac{\lambda^2-\lambda^{-2}}{2} \end{array}\right]\Phi_l.\label{lax:sdmKdV}
\end{equation}
Apparently, the dispersion relation suggested from the linear problem is different from the one in
Theorem \ref{thm:solutions}. However, putting
\begin{equation}
 p_i = \frac{1}{\epsilon}\frac{\lambda_i^2-1}{\lambda_i^2+1}
\end{equation}
in (\ref{tau:entries}), then $f_n^{(i)}$ can be rewritten as
\begin{align*}
 f_n^{(i)}&\Bumpeq
\alpha_i\left(\frac{1}{\epsilon}\frac{\lambda_i^2-1}{\lambda_i^2+1}\right)^n
\lambda_i^l
e^{\frac{1}{2}(\lambda_i^2-\lambda_i^{-2})s'+\frac{\lambda_i^2+1}{\lambda_i^2-1}\epsilon y}
+ \beta_i\left(-\frac{1}{\epsilon}\frac{\lambda_i^2-1}{\lambda_i^2+1}\right)^n
\lambda_i^{-l}
e^{-\frac{1}{2}(\lambda_i^2-\lambda_i^{-2})s'+\frac{\lambda_i^2+1}{\lambda_i^2-1}\epsilon y},
\end{align*}
in which the dispersion relation with respect to $l$ and $s'$ is consistent with
(\ref{lax:sdmKdV}). We have chosen the parametrization as in (\ref{tau:entries}) so that the
continuous limits explained in Section \ref{sec:continuous_limit} become simpler.
\end{remark}
%
\section{B\"acklund Transformations}
In this section we discuss the B\"acklund transformation of the continuous motion of plane discrete
curves. The B\"acklund transformation of the plane discrete curves has already been formulated in \cite{IKMO:discrete_curve}:
\begin{proposition}\label{prop:BT0}
 Let $\gamma_l$ be a discrete curve of segment length $\epsilon$.
Let  $\theta_l$ be the potential function  defined by
\begin{equation}
 \frac{\gamma_{l+1}-\gamma_n}{\epsilon}
= \left[\begin{array}{c}\cos\psi_l \\\sin\psi_l\end{array}\right],\quad
\psi_l = \frac{\theta_{l+1}+\theta_{l}}{2}.\label{discrete_angle2}
\end{equation}
For a nonzero constant $\lambda$, take a solution $\widetilde\theta_n$ to the following equation
\begin{equation}
 \tan\left(\frac{\widetilde\theta_{l+1} - \theta_l}{4}\right)=\frac{\frac{1}{\lambda}+\epsilon}{\frac{1}{\lambda}-\epsilon}\tan\left(\frac{\widetilde\theta_l-\theta_{l+1}}{4}\right),\label{BT0:dpmKdV}
\end{equation}
then
\begin{equation}
 \widetilde\gamma_l
=\gamma_l  +\frac{1}{\lambda}~R\left(\frac{\widetilde\theta_l-\theta_{l+1}}{2}\right)~
\frac{\gamma_{l+1}-\gamma_l}{\epsilon}\label{BTn:gamma}
\end{equation}
is a discrete curve with the potential function $\widetilde\theta_l$.
\end{proposition}
We next extend the B\"acklund transformation to that of motion of discrete curves. In order to do
so, we first present the B\"acklund transformation to the semi-discrete potential mKdV equation:
%
\begin{lemma}\label{lem:BT_dpmKdV}
Let $\theta_l$ be a solution to the semi-discrete potential mKdV equation (\ref{sdpmKdV}). A
 function $\widetilde\theta_l$ satisfying the following system of equations
\begin{align}
& \left(\frac{1}{\lambda}-\epsilon\right) \tan\frac{\widetilde\theta_{l+1} - \theta_l}{4}=\left(\frac{1}{\lambda}+\epsilon\right)\tan\frac{\widetilde\theta_l-\theta_{l+1}}{4},\label{BT1:sdpmKdV}\\[2mm]
&
 \left(\frac{1}{\lambda}+\epsilon\right)\frac{\widetilde\theta_{l}^{\,\prime}}{4\cos^2\dfrac{\widetilde\theta_{l}-\theta_{l+1}}{4}}
+\left(\frac{1}{\lambda}-\epsilon\right)\frac{\theta_{l}^{\,\prime}}{4\cos^2\dfrac{\widetilde\theta_{l+1}-\theta_{l}}{4}}
= \tan\frac{\widetilde\theta_{l}-\theta_{l+1}}{4}
+ \tan\frac{\widetilde\theta_{l+1}-\theta_{l}}{4},
\label{BT2:sdpmKdV}
\end{align}
gives another solution to eq.(\ref{sdpmKdV}). We call $\widetilde\theta_l$ a B\"acklund
 transform of $\theta_l$.
\end{lemma}
\noindent\textbf{Proof.} First compute addition of
(\ref{BT2:sdpmKdV})$_{l-1}$ and the derivative of
(\ref{BT1:sdpmKdV})$_{l-1}$. Then, by using (\ref{BT1:sdpmKdV}),
eliminate $\lambda$ from this equation and (\ref{BT2:sdpmKdV}) respectively.
Adding those two equations yields
\begin{equation}
\begin{split}
&
 \left(\frac{\epsilon}{2}\,\widetilde\theta_l^{\,\prime}\cos\frac{\widetilde\theta_{l+1}-\widetilde\theta_{l-1}}{4}-\sin\frac{\widetilde\theta_{l+1}-\widetilde\theta_{l-1}}{4}\right)
 \sin\frac{\widetilde\theta_{l+1}+\widetilde\theta_{l-1}-2\theta_l}{4}\\[2mm]
&=
\left(\frac{\epsilon}{2}\,\theta_l^{\,\prime}\cos\frac{\theta_{l+1}-\theta_{l-1}}{4}-\sin\frac{\theta_{l+1}-\theta_{l-1}}{4}\right)
\sin\frac{\theta_{l+1}+\theta_{l-1}-2\widetilde\theta_l}{4},
\end{split}
\end{equation}
which implies Lemma \ref{lem:BT_dpmKdV}.\qed
%
\begin{proposition}\label{prop:BT_discrete}
 Let $\gamma_l$ be a motion of discrete curve. Take a B\"acklund transform
$\widetilde\theta_l$ of $\theta_l$ defined in Lemma \ref{lem:BT_dpmKdV}. Then
\begin{equation}
 \widetilde\gamma_l = \gamma_l + \frac{1}{\lambda}~
R\left(\frac{\widetilde\theta_l -\theta_{l+1}}{2}\right)
~\frac{\gamma_{l+1}-\gamma_l}{\epsilon}
\label{BT_discrete}
\end{equation}
is a motion of discrete curve with potential function $\widetilde\theta_l.$ We call
 $\widetilde{\gamma}_l$ a B\"acklund transform of $\gamma_l$.
\end{proposition}

\noindent{\bf Proof.} It suffices to show that $\widetilde\gamma_l$ satisfies
eqs.(\ref{iso0}), (\ref{discrete_FS}) and (\ref{gamma_s}) with potential function
$\widetilde\theta_l$, but eqs.(\ref{iso0}) and (\ref{discrete_FS}) follow from Proposition
\ref{prop:BT0} immediately.
Because the system \eqref{BT1:sdpmKdV}--\eqref{BT2:sdpmKdV} yields
\begin{equation*}
\left(1-\sqrt{-1}\frac{\epsilon}{2}\,\widetilde{\theta}_l^{\,\prime}\right)
e^{\sqrt{-1}\,\frac{\widetilde{\theta}_{l+1}-\theta_l}{2}}
=
\left(1-\sqrt{-1}\frac{\epsilon}{2}\theta_l^{\,\prime}\right)
e^{\sqrt{-1}\,\frac{\widetilde{\theta}_l-\theta_{l+1}}{2}}
+\frac{\sqrt{-1}}{\lambda}\frac{\widetilde{\theta}_l^{\,\prime}+\theta_l^{\,\prime}}{2},
\end{equation*}
we identify $\mathbb{R}^2$ with $\mathbb{C}$,
so that the motion $\widetilde{\gamma}_l$ satisfies
\begin{equation*}
\widetilde{\gamma}_l^{\,\prime}
=e^{\sqrt{-1}\,\frac{\widetilde{\theta}_{l+1}+\widetilde{\theta}_l}{2}}\left(1-\sqrt{-1}\frac{\epsilon}{2}\,\widetilde{\theta}_l^{\,\prime}\right)\widetilde\gamma_l
=\frac{\widetilde{\gamma}_{l+1}-\widetilde{\gamma}_l}{\epsilon}\left(1-\sqrt{-1}\tan\frac{\widetilde{\theta}_{l+1}-\widetilde{\theta}_{l-1}}{4}\right),
\end{equation*}
which implies \eqref{gamma_s} with $2\widetilde{K}_l=\widetilde{\theta}_{l+1}-\widetilde{\theta}_{l-1}$.
\qed
\begin{remark}\rm
In \cite{Hoffmann:LN,HK}, the B\"acklund transformation of the motions of discrete plane curves
 described in this paper is characterized by the cross ratio of the four points $\gamma_l$,
 $\gamma_{l+1}$, $\widetilde\gamma_l$ and $\widetilde\gamma_{l+1}$ being constant. In fact, we can
 verify by direct computation that for the B\"acklund transformation given in Proposition
 \ref{prop:BT_discrete}, the cross ratio of those four points is $-\lambda^2\epsilon^2$.  In case of
 continuous motions of smooth plane curves, Darboux matrices of the B\"acklund transformations are
 obtained in \cite{Sym}. Sym gave explicit formulas for smooth plane curves derived from
 multi-soliton solutions to the mKdV equation via the iterated B\"acklund transformations.
\end{remark}
\section{Continuous Limits}\label{sec:continuous_limit}
In \cite{IKMO:discrete_curve}, the discrete motion of discrete plane curves and the continuous
motion of smooth plane curves have been formulated, together with the B\"acklund transformations and
the explicit formulas in terms of the $\tau$ functions. They are described by the discrete potential
modified KdV equation and the potential modified KdV equation, respectively.  In this section, we
present the two continuous limits: one from the discrete motion of discrete plane curves to their
continuous motion discussed in the preceding sections, another one from the continuous motion of
discrete plane curves to the continuous motion of smooth plane curves.

We first summarize the formulations of three kinds of curve 
motions and explicit solutions.
For convenience, we identify Euclidean plane 
$\mathbb{R}^2$ with complex plane $\mathbb{C}$.

\par\bigskip

\noindent \textbf{(1) Discrete motion of discrete plane curves.}\\[2mm]
\noindent Motion of curves:
\begin{align}
 &\left|\frac{\gamma_{n+1}^m-\gamma_n^m}{a_n}\right|=1,\label{discrete:gamma:arc-length}\\
 & \frac{\gamma_{n+1}^m-\gamma_n^m}{a_n}=e^{\sqrt{-1}K_n^m}~\frac{\gamma_{n}^m-\gamma_{n-1}^m}{a_{n-1}},\label{discrete:gamma:n}\\
 & \frac{\gamma_{n}^{m+1}-\gamma_n^m}{b_m}=e^{\sqrt{-1}W_n^m}~\frac{\gamma_{n+1}^m-\gamma_{n}^m}{a_{n}}.\label{discrete:gamma:m}
\end{align}
Here, $n,m\in\mathbb{Z}$ denote the discrete independent variables corresponding to space and time,
respectively. Moreover, $a_n$, $b_m$ are real arbitrary functions of the indicated variables, which
correspond to the segment length of the curves and time interval, respectively.
\par\bigskip
\noindent Potential function:
\begin{equation}
 K_n^m = \frac{\theta_{n+1}^m-\theta_{n-1}^m}{2},\quad 
 W_n^m = \frac{\theta_{n}^{m+1}-\theta_{n+1}^m}{2}.\label{discrete:potential}
\end{equation}
\noindent Compatibility condition:
\begin{equation}
 \tan\left(\frac{\theta_{n+1}^{m+1}-\theta_n^m}{4}\right)=\frac{b_m+a_n}{b_m-a_n} \tan\left(\frac{\theta_{n}^{m+1}-\theta_{n+1}^m}{4}\right).\label{eqn:dpmKdV}
\end{equation}
\noindent Explicit formula in terms of $\tau$ function:
\begin{equation}
 \theta_n^m = \frac{2}{\sqrt{-1}}\log\frac{\tau_n^m}{\taus^n_m},\quad 
\gamma_n^m = \left[\begin{array}{c}{\displaystyle -\frac{1}{2}\left(\log\tau_n^m\taus_n^m\right)_y} \\[2mm]
{\displaystyle  \frac{1}{2\sqrt{-1}}\left(\log\frac{\tau_n^m}{\taus^m_n}\right)_y}\end{array}\right].\label{discrete:rep_formula}
\end{equation}
\noindent Soliton type solutions:
\begin{equation}
 \tau_n^m = \exp\left[-\left(\sum_{n'}^{n-1}a_{n'} + \sum_{m'}^{m-1}b_{m'}\right)y\right]
~\det\left(f_{j-1}^{(i)}\right)_{i,j=1,\ldots,N},
\label{discrete:Casorati} 
\end{equation}
\begin{equation}
f_k^{(i)}= \alpha_ip_i^k \prod_{n'}^{n-1}(1-a_{n'}p_i)^{-1}\prod_{m'}^{m-1}(1-b_{m'}p_i)^{-1}e^{\frac{1}{p_i}y }+
\beta_i(-p_i)^k \prod_{n'}^{n-1}(1+a_{n'}p_i)^{-1}\prod_{m'}^{m-1}(1+b_{m'}p_i)^{-1}e^{-  \frac{1}{p_i}y }.
\label{discrete:Casorati_entries}
\end{equation}
\noindent \textbf{(2) Continuous motion of discrete plane curves.}\\[2mm]
\noindent Motion of curves:
\begin{align}
 &\left|\frac{\gamma_{l+1}-\gamma_l}{\epsilon}\right|=1,\label{semi-discrete:gamma:arc-length}\\
 & \frac{\gamma_{l+1}-\gamma_l}{\epsilon}=e^{\sqrt{-1}K_l}~\frac{\gamma_{l}-\gamma_{l-1}}{\epsilon},\label{semi-discrete:gamma:l}\\
 & \frac{d\gamma_l}{ds}=\frac{e^{-\sqrt{-1}\frac{K_l}{2}}}{\cos\frac{K_l}{2}}~
\frac{\gamma_{l+1}-\gamma_{l}}{\epsilon}.\label{semi-discrete:gamma:s}
\end{align}
\noindent Potential function:
\begin{equation}
 K_l =  \frac{\theta_{l+1}-\theta_{l-1}}{2}.\label{semi-discrete:potential}
\end{equation}
\noindent Compatibility condition:
\begin{equation}
\frac{d\theta_l}{ds} = \frac{2}{\epsilon}\tan\left(\frac{\theta_{l+1}-\theta_{l-1}}{4}\right). \label{eqn:sdpmKdV}
\end{equation}
\noindent Explicit formula in terms of $\tau$ function:
\begin{equation}
 \theta_l = \frac{2}{\sqrt{-1}}\log\frac{\tau_l}{\tau^*_l},\quad 
\gamma_l = \left[\begin{array}{c}{\displaystyle -\frac{1}{2}\left(\log\tau_l\tau^*_l\right)_y} \\[2mm]
{\displaystyle  \frac{1}{2\sqrt{-1}}\left(\log\frac{\tau_l}{\tau^*_l}\right)_y}\end{array}\right].\label{semi-discrete:rep_formula}
\end{equation}
\noindent Soliton type solutions:
\begin{equation}
 \tau_l = \exp\left[-\left(s+\epsilon l\right)y\right]
~\det\left(f_{j-1}^{(i)}\right)_{i,j=1,\ldots,N},
\label{semi-discrete:Casorati} 
\end{equation}
\begin{equation}
 f_k^{(i)} = \alpha_i p_i^k (1-\epsilon p_i)^{-l} e^{\frac{p_i}{1-\epsilon^2p_i^2}s + \frac{1}{p_i}y}
+\beta_i (-p_i)^k (1+\epsilon p_i)^{-l} e^{-\frac{p_i}{1-\epsilon^2p_i^2}s - \frac{1}{p_i}y}.
\label{semi-discrete:Casorati_entries}
\end{equation}
\noindent \textbf{(3) Continuous motion of smooth plane curves.}\\[2mm]
\noindent Motion of curves: 
\begin{align}
 &\left|\gamma'\right|=1,\label{continuous:arc-length}\\
 &\frac{\partial}{\partial x}\gamma' = \sqrt{-1}\kappa~\gamma',\label{continuous:gamma:x}\\
 & \frac{\partial}{\partial t}\gamma'=
 -\sqrt{-1}\left(\kappa''+\frac{\kappa^3}{2}\right)~\gamma'.\label{continuous:gamma:t}
\end{align}
Here $\gamma=\gamma(x,t)\in\mathbb{R}^2\simeq \mathbb{C}$ is arc-length parametrized curve, $x$ and $t$
denote arc-length and time, respectively, and $'=\partial_x$. Moreover, $\kappa=\kappa(x,t)$ is the curvature.
\par\bigskip
\noindent Potential function:
\begin{equation}
 \kappa=\theta'.
\end{equation}
\noindent Compatibility condition:
\begin{equation}
 \theta_t + \frac{1}{2}(\theta_x)^3+\theta_{xxx}=0.\label{eqn:pmKdV}
\end{equation}
\noindent Explicit formula in terms of $\tau$ function:
\begin{equation}
 \theta = \frac{2}{\sqrt{-1}}\log\frac{\tau}{\taus},\quad 
\gamma = \left[\begin{array}{c}{\displaystyle -\frac{1}{2}\left(\log\tau\taus\right)_y} \\[2mm]
{\displaystyle  \frac{1}{2\sqrt{-1}}\left(\log\frac{\tau}{\taus}\right)_y}\end{array}\right].\label{continuous:rep_formula}
\end{equation}
\noindent Soliton type solutions:
\begin{equation}
 \tau = e^{-xy}~\det\left(f_{j-1}^{(i)}\right)_{i,j=1,\ldots,N},\label{continuous:Casorati} 
\end{equation}
\begin{equation}
f_k^{(i)}= \alpha_ip_i^k e^{p_i x - 4p_i^3t+\frac{1}{p_i}y }+
\beta_i(-p_i)^k e^{-p_i x + 4p_i^3t -  \frac{1}{p_i}y }.
\label{continuous:Casorati_entries}
\end{equation}
\begin{theorem}\label{thm:continuous_limit}\hfill
\begin{enumerate}
 \item Putting 
\begin{equation}
\begin{array}{c}
{\displaystyle a_n=a\ (\text{const.}),\quad b_m=b\ (\text{const.}),\quad \delta=\frac{a+b}{2},\quad \epsilon=\frac{a-b}{2},} \\[2mm]
{\displaystyle \frac{s}{\delta}=n+m,\quad l = n-m,}
\end{array}
\label{discrete_to_semidiscrete:parametrization}
\end{equation}
and taking the limit $\delta\to 0$, the discrete motion of discrete plane curves yields the
       continuous motion of discrete plane curves.
\item Putting 
\begin{equation}
  x = \epsilon l + s,\quad t = -\frac{\epsilon^2}{6}s,\label{continuous:parametrization}
\end{equation}
and taking the limit $\epsilon\to 0$, the continuous motion of discrete plane curves yields the
continuous motion of smooth plane curves.
\end{enumerate}
\end{theorem}

Theorem \ref{thm:continuous_limit} can be verified by tedious but straightforward calculations. In
fact, the statement (1) can be checked by substituting the parametrization
(\ref{discrete_to_semidiscrete:parametrization}) into
(\ref{discrete:gamma:arc-length})--(\ref{discrete:Casorati_entries}), expanding in terms of powers
of $\delta$ and taking the limit $\delta\to 0$. The statement (2) is also checked by a similar
manner. We note that the limiting procedures presented in
(\ref{discrete_to_semidiscrete:parametrization}) and (\ref{continuous:parametrization})
have been obtained in \cite{Hirota:semi-discrete_mKdV} and \cite{Hirota:dpmKdV} on the level of the
equations for $\theta$. Theorem \ref{thm:continuous_limit} claims that the procedure applies to the
curve motions and solutions. Also, it should be noted that limiting procedure also applies to the
B\"acklund transformations.

In order to demonstrate the calculation, we here discuss the limits of the $\tau$ functions
corresponding to the soliton type solutions. Substituting
(\ref{discrete_to_semidiscrete:parametrization}) into (\ref{discrete:Casorati_entries}), we have
\begin{align*}
& (1-ap_i)^{-n}(1-bp_i)^{-m} = (1-ap_i)^{-\frac{1}{2}\left(\frac{s}{\delta}+l\right)}(1-bp_i)^{-\frac{1}{2}\left(\frac{s}{\delta}-l\right)}
= e^{-\frac{s}{2\delta}\log\left[1-2\delta p_i+(\delta^2-\epsilon^2)p_i^2\right]}
\left(\frac{1-\epsilon p_i-\delta p_i}{1+\epsilon p_i-\delta p_i}\right)^{-\frac{l}{2}}.
\end{align*}
Noticing that
\begin{align*}
 &\log\left[1-2\delta p_i+(\delta^2-\epsilon^2)p_i^2\right]
= \log\omega_i - \frac{2p_i}{\omega_i}\delta+O(\delta^2),\quad \omega_i = 1- \epsilon^2p_i^2,
\end{align*}
we get
\begin{displaymath}
 (1-ap_i)^{-n}(1-bp_i)^{-m} \sim e^{-\frac{\log\omega_i}{2\delta}s}\times e^{\frac{p_i}{\omega_i}s}~
\left(\frac{1-\epsilon p_i}{1+\epsilon p_i}\right)^{-\frac{l}{2}}.
\end{displaymath}
Similarly, we have
\begin{displaymath}
 (1+ap_i)^{-n}(1+bp_i)^{-m} \sim e^{-\frac{\log\omega_i}{2\delta}s}\times e^{-\frac{p_i}{\omega_i}s}~
\left(\frac{1+\epsilon p_i}{1-\epsilon p_i}\right)^{-\frac{l}{2}}.
\end{displaymath}
Therefore $f_k^{(i)}$ yields 
\begin{align}
f_k^{(i)}&\sim \alpha_ip_i^k e^{-\frac{\log\omega_i}{2\delta}s}~ e^{\frac{p_i}{\omega_i}s}~\left(\frac{1-\epsilon p_i}{1+\epsilon p_i}\right)^{-\frac{l}{2}}e^{\frac{1}{p_i}y }
+ \beta_i(-p_i)^k e^{-\frac{\log\omega_i}{2\delta}s}~
 e^{-\frac{p_i}{\omega_i}s}~\left(\frac{1+\epsilon p_i}{1-\epsilon
 p_i}\right)^{-\frac{l}{2}}e^{-\frac{1}{p_i}y } \notag \\[2mm]
&=e^{-\frac{\log\omega_i}{2\delta}s}(1-\epsilon^2p_i^2)^{\frac{l}{2}}
\left[\alpha_ip_i^k \left(1-\epsilon p_i\right)^{-l}e^{\frac{p_i}{\omega_i}s+\frac{1}{p_i}y}
+ \beta_i(-p_i)^k \left(1+\epsilon p_i\right)^{-l}e^{-\frac{p_i}{\omega_i}s-\frac{1}{p_i}y}\right] ,\label{entries:limit1}
\end{align}
as $\delta\sim 0$. The prefactors of the entries in (\ref{entries:limit1}) can be factored out of
the determinant, and it is easily seen that the overall factor does not affect the solutions,
namely, if we remove overall factor from the $\tau$ functions, it gives the same $\theta$ and
$\gamma$, as seen from (\ref{discrete:rep_formula}). This implies that the determinant in
(\ref{discrete:Casorati}) yields that in (\ref{semi-discrete:Casorati}) up to this trivial
multiplicative factor.  Also, the exponential factor in (\ref{discrete:rep_formula}) becomes that in
(\ref{semi-discrete:rep_formula}) under the parametrization
(\ref{discrete_to_semidiscrete:parametrization}).  Therefore, we have shown that
(\ref{discrete:Casorati}) is reduced to (\ref{semi-discrete:Casorati}) as $\delta\to 0$.

Similarly, substituting (\ref{continuous:parametrization}) into
(\ref{semi-discrete:Casorati_entries}), we have
\begin{align*}
& \left(1-\epsilon p_i\right)^{-l}e^{\frac{p_i}{1-\epsilon^2p_i^2}s}
=\exp\left[-\left(\frac{x}{\epsilon}+\frac{6t}{\epsilon^3}\right)\log\left(1-\epsilon p_i\right) 
+ \frac{p_i}{1-\epsilon^2p_i^2}s\right]
 = \exp\left[\frac{3p_i^2}{\epsilon}t+\left(p_i x-4p_i^3t\right)+O(\epsilon)\right],
\end{align*}
and
\begin{displaymath}
 \left(1+\epsilon p_i\right)^{-l}e^{-\frac{p_i}{1-\epsilon^2p_i^2}s}=\exp\left[\frac{3p_i^2}{\epsilon}t-\left(p_i x-4p_i^3t\right)+O(\epsilon)\right],
\end{displaymath}
from which we obtain as $\epsilon\sim 0$
\begin{align}
 f_k^{(i)}
\sim e^{\frac{3p_i^2}{\epsilon}t}~\left[\alpha_ip_i^k e^{ p_i x-4p_i^3t +\frac{1}{p_i}y}
+\beta_i(-p_i)^k e^{ -p_i x+4p_i^3t -\frac{1}{p_i}y}\right].\label{entries:limit2}
\end{align}
The prefactor in (\ref{entries:limit2}) does not affect the solutions. Also, the exponential factor
in (\ref{semi-discrete:rep_formula}) becomes that in (\ref{continuous:rep_formula})
under the parametrization (\ref{continuous:parametrization}).  Therefore, we have
shown that (\ref{semi-discrete:Casorati}) is reduced to (\ref{continuous:Casorati}) as $\epsilon\to 0$.
\appendix
\makeatletter
\@addtoreset{equation}{section}
\renewcommand{\theequation}{\@Alph\c@section.\@arabic\c@equation}
\makeatother
\section{Derivation of bilinear equations (\ref{bl5}) and (\ref{bl8})}
In this appendix we prove Proposition \ref{prop:generic}. As mentioned in Section \ref{sec:tau}, it
suffices to show that the $\tau$ function given in (\ref{sigma}) and (\ref{sigma:entries}) actually
satisfies the bilinear equations (\ref{bl5}) and (\ref{bl8}).
\subsection{Equation (\ref{bl5})}
We define the $\tau$ function $\sigma^k_{l,m}(u,v;y)$ by
\begin{equation}
  \sigma^k_{l,m}(u,v;y) = \det\left(f_{k+j-1}^{(i)}(l,m)\right)_{i,j=1,\ldots,N},\label{sigma2}
\end{equation}
where the entries of determinant satisfy the linear relations
\begin{align}
& \frac{f_k^{(i)}(l,m)-f_k^{(i)}(l-1,m)}{a}=f_{k+1}^{(i)}(l,m),\quad
 \frac{f_k^{(i)}(l,m)-f_k^{(i)}(l,m-1)}{b}=f_{k+1}^{(i)}(l,m), \label{sigma:linear1}\\[2mm]
& \partial_u  f_k^{(i)}(l,m)=f_k^{(i)}(l+1,m),\quad
\partial_v  f_k^{(i)}(l,m)=f_k^{(i)}(l,m+1),\quad
\partial_y  f_k^{(i)}(l,m)=f_{k-1}^{(i)}(l,m).\label{sigma:linear2}
\end{align}
Note that $f_k^{(i)}(l,m)$ given in (\ref{sigma:entries}) satisfy the above relations. In order to
prove (\ref{bl5}), it is convenient to consider $\rho^k_{l,m}$ defined by
\begin{equation}
 \rho^k_{l,m}(u,v;y) = \det\left(f_{k}^{(i)}(l-j+1,m)\right)_{i,j=1,\ldots,N},\label{rho}
\end{equation}
instead of $\sigma^k_{l,m}$. Here, $\sigma^k_{l,m}$ and $\rho^k_{l,m}$ are related as
\begin{equation}
 \rho^k_{l,m} = (-a)^{N(N-1)/2}~\sigma^k_{l,m},\label{rho_and_sigma} 
\end{equation}
which can be easily verified by manipulating the columns of determinant with the first equation in
(\ref{sigma:linear1}). We also introduce a notation
\begin{equation}
 \rho^k_{l,m} = \left|~ \bm{0}^k_m,\bm{1}^k_m,\cdots,\bm{N-2}^k_m,\bm{N-1}^k_m\right|,\quad
\bm{j}^k_m=\left[\begin{array}{c} f_k^{(1)}(l-j,m)\\[2mm]f_k^{(2)}(l-j,m)\\[2mm]\vdots \\[2mm]f_k^{(N)}(l-j,m)\end{array}\right].
\end{equation}
It is possible to reduce (\ref{bl5}) to one of the Pl\"ucker relations which are quadratic quadratic
identities of determinants whose columns are appropriately shifted.  To this end, we construct such
formulas that express the determinants in the Pl\"ucker relations in terms of derivative or shift of
discrete variable of $\rho^k_{l,m}(u,v;y)$ by using the linear relations of the entries. For 
details of the technique, we refer to \cite{Hirota:book,OKMS:RT,OHTI:dKP,MKO:dRT,MO:dNLS_dark}.
\begin{lemma}\label{lem:appendix_differential_formula}
 The following formulas hold:
\begin{align}
 &\partial_u~\rho^k_{l,m}=\left|~ \bm{-1},\bm{1},\cdots,\bm{N-2},\bm{N-1}\right|,\label{dif1}\\[2mm]
 &\rho^{k-1}_{l,m}=a^{N-1}~\left|~ \bm{0},\bm{1},\cdots,\bm{N-2},\bm{N-1}^{k-1}\right|,\label{dif2}\\[2mm]
 &\rho^{k-1}_{l,m}=a^{N-1}~\left|~ \bm{0},\bm{1},\cdots,\bm{N-2},\bm{N-2}^{k-1}\right|,\label{dif3}\\[2mm]
 &\left(\partial_u-1\right)~\rho^{k-1}_{l,m}=a^{N-1}~\left|~ \bm{-1},\bm{1},\cdots,\bm{N-2},\bm{N-1}^{k-1}\right|.\label{dif4}
\end{align}
Note that the superscripts of column vectors are shown only when $k$ is shifted for notational simplicity.
\end{lemma}
\noindent{\bf Proof.}\quad Equation (\ref{dif1}) follows from the differential rule of determinants and the
fist equation of (\ref{sigma:linear1}). Next, applying the first equation of the difference rule
(\ref{sigma:linear2}) to the first column of $\rho^{k-1}_{l,m}$, we have
\begin{displaymath}
 \rho^{k-1}_{l,m}=\left|~ \bm{0}^{k-1},\bm{1}^{k-1},\cdots,\bm{N-1}^{k-1}\right|
=\left|~ \bm{0}^{k-1}-\bm{1}^{k-1},\bm{1}^{k-1},\cdots,\bm{N-1}^{k-1}\right|
=a\left|~ \bm{0}^{k},\bm{1}^{k-1},\cdots,\bm{N-1}^{k-1}\right|.
\end{displaymath}
Repeating this procedure for the $j$-th column ($j=2,3,\ldots,N-1$), we get
\begin{displaymath}
  \rho^{k-1}_{l,m}=a^{N-1}~\left|~ \bm{0}^{k},\bm{1}^{k},\cdots,\bm{N-2}^{k},\bm{N-1}^{k-1}\right|,
\end{displaymath}
which is (\ref{dif2}). Applying (\ref{sigma:linear1}) to the $N$-th column of (\ref{dif2}), we
obtain (\ref{dif3}).

Finally, differentiating (\ref{dif3}) by $u$ yields
\begin{align*}
 \partial_u~\rho^{k-1}_{l,m}&=a^{N-1}~\left|~ \bm{-1}^{k},\bm{1}^{k},\cdots,\bm{N-2}^{k},\bm{N-1}^{k-1}\right|
+ a^{N-1}~\left|~ \bm{0}^{k},\bm{1}^{k},\cdots,\bm{N-2}^{k},\bm{N-2}^{k-1}\right|\\[2mm]
&= a^{N-1}~\left|~ \bm{-1}^{k},\bm{1}^{k},\cdots,\bm{N-2}^{k},\bm{N-1}^{k-1}\right| + \rho^{k-1}_{l,m}
\end{align*}
which is equivalent to (\ref{dif4}). Thus we have proved Lemma \ref{lem:appendix_differential_formula}.\qed

\par\bigskip

Now consider the Pl\"ucker relation (see, for example, \cite{OKMS:RT})
\begin{equation}
\begin{split}
0 &= \left|~ \bm{-1},\bm{0},\bm{1},\cdots,\bm{N-2}\right|\times
\left|~ \bm{1},\cdots,\bm{N-2},\bm{N-1},\bm{N-1}^{k-1}\right|\\
&+ \left|~ \bm{0},\bm{1},\cdots,\bm{N-2},\bm{N-1}\right|\times
\left|~ \bm{-1},\bm{1},\cdots,\bm{N-2},\bm{N-1}^{k-1}\right|\\
&- \left|~ \bm{0},\bm{1},\cdots,\bm{N-2},\bm{N-1}^{k-1}\right|\times
\left|~ \bm{-1},\bm{1},\cdots,\bm{N-2},\bm{N-1}\right|.
\end{split}\label{pl1}
\end{equation}
(\ref{pl1}) is rewritten by using Lemma \ref{lem:appendix_differential_formula} as
\begin{align}
0&=\rho^{k}_{l+1,m}\times a^{-(N-1)}\rho^{k-1}_{l-1,m} + \rho^{k}_{l,m}\times a^{-(N-1)}(\partial_u-1)~\rho^{k-1}_{l,m}
- a^{-(N-1)}\rho^{k-1}_{l,m}\times\partial_u~\rho^{k}_{l,m}\notag\\
&= a^{-(N-1)}\left[\left(D_u-1\right)~\rho^{k-1}_{l,m}\cdot\rho^{k}_{l,m}+\rho^{k}_{l+1,m}\rho^{k-1}_{l-1,m}\right],
\end{align}
which implies (\ref{bl5}).

\subsection{Equation (\ref{bl8})}
We derive (\ref{bl8}) from (\ref{bl5}) and (\ref{bl10}). We first introduce $F^k_{l,m}$ by
the subtraction of the right hand side of (\ref{bl8}) from the left hand side
\begin{equation}
 F^k_{l,m}:=\frac{1}{2}D_uD_y~\sigma^{k}_{l,m}\cdot\sigma^{k}_{l,m} - a(\sigma^{k}_{l,m})^2 
+ a\sigma^{k+1}_{l+1,m}\sigma^{k-1}_{l-1,m},
\end{equation}
and consider
\begin{align}
 P&:=F^k_{l,m}\left(\sigma^{k-1}_{l,m}\right)^2 - F^{k-1}_{l,m}\left(\sigma^{k}_{l,m}\right)^2\notag\\
&=\left[\frac{1}{2}D_uD_y~\sigma^{k}_{l,m}\cdot\sigma^{k}_{l,m} - a(\sigma^{k}_{l,m})^2 
+ a\sigma^{k+1}_{l+1,m}\sigma^{k-1}_{l-1,m}\right]~\left(\sigma^{k-1}_{l,m}\right)^2\notag\\
&- \left(\sigma^{k}_{l,m}\right)^2~\left[\frac{1}{2}D_uD_y~\sigma^{k-1}_{l,m}\cdot\sigma^{k-1}_{l,m} -
 a(\sigma^{k-1}_{l,m})^2 + a\sigma^{k}_{l+1,m}\sigma^{k-2}_{l-1,m}\right].\label{P}
\end{align}
Equation (\ref{P}) can be rewritten as
\begin{equation}
 P=D_y\left(D_x~\sigma^{k}_{l,m}\cdot\sigma^{k-1}_{l,m}\right)\cdot \sigma^{k}_{l,m}\sigma^{k-1}_{l,m}
+ a\sigma^{k+1}_{l+1,m}\sigma^{k-1}_{l-1,m}\sigma^{k-1}_{l,m}\sigma^{k-1}_{l,m}
- a\sigma^{k}_{l+1,m}\sigma^{k-2}_{l-1,m}\sigma^{k}_{l,m}\sigma^{k}_{l,m},\label{bl8:derivation}
\end{equation}
where we have used the exchange formula of the $D$-operator \cite{Hirota:book}
\begin{equation}
 \left(D_uD_y~f\cdot f\right)g^2 -  f^2\left(D_uD_y~g\cdot g\right)=2D_y~\left(D_u~f\cdot g\right)\cdot fg,
\end{equation}
for arbitrary functions $f$ and $g$.
We manipulate the first term of (\ref{bl8:derivation}) as follows. Using (\ref{bl5}) and noticing
$D_y~f\cdot f=0$, we have
\begin{displaymath}
 D_y\left(D_x~\sigma^{k}_{l,m}\cdot\sigma^{k-1}_{l,m}\right)\cdot \sigma^{k}_{l,m}\sigma^{k-1}_{l,m}
=  D_y\left(-\sigma^{k-1}_{l,m}\sigma^{k}_{l,m}+\sigma^{k}_{l+1,m}\sigma^{k-1}_{l-1,m}\right)\cdot
 \sigma^{k}_{l,m}\sigma^{k-1}_{l,m}
= D_y~\sigma^{k}_{l+1,m}\sigma^{k-1}_{l-1,m}\cdot \sigma^{k}_{l,m}\sigma^{k-1}_{l,m}.
\end{displaymath}
Then applying another exchange formula
\begin{equation}
 D_y~\alpha\beta\cdot \gamma\delta = \left(D_y~\alpha\cdot \gamma\right)\beta\delta + \left(D_y~\beta\cdot \delta\right)\alpha\gamma,
\end{equation}
for arbitrary functions $\alpha$, $\beta$, $\gamma$, $\delta$, we get
\begin{align*}
& D_y~\sigma^{k}_{l+1,m}\sigma^{k-1}_{l-1,m}\cdot \sigma^{k}_{l,m}\sigma^{k-1}_{l,m}
= \left(D_y~\sigma^{k}_{l+1,m}\cdot\sigma^{k}_{l,m}\right)\sigma^{k-1}_{l-1,m}\sigma^{k-1}_{l,m}
+ \left(D_y~\sigma^{k-1}_{l-1,m}\cdot\sigma^{k-1}_{l,m}\right)\sigma^{k}_{l+1,m}\sigma^{k}_{l,m}\\[2mm]
&=
\left(\sigma^{k}_{l+1,m}\sigma^{k}_{l,m} -a \sigma^{k+1}_{l+1,m}\sigma^{k-1}_{l,m}\right)\sigma^{k-1}_{l-1,m}\sigma^{k-1}_{l,m}
+ \left(-\sigma^{k-1}_{l,m}\sigma^{k-1}_{l-1,m} 
+ a \sigma^{k}_{l,m}\sigma^{k-2}_{l-1,m}\right)\sigma^{k}_{l+1,m}\sigma^{k}_{l,m}\\[2mm]
& = -a \sigma^{k+1}_{l+1,m}\sigma^{k-1}_{l,m}\sigma^{k-1}_{l-1,m}\sigma^{k-1}_{l,m}
+ a \sigma^{k}_{l,m}\sigma^{k-2}_{l-1,m}\sigma^{k}_{l+1,m}\sigma^{k}_{l,m}
\end{align*}
where we have used (\ref{bl10}) in the second equality. Substituting the above result into
(\ref{bl8:derivation}), we see that $P=0$. Therefore, it follows from (\ref{P}) that
\begin{equation}
 \frac{1}{2}D_uD_y~\sigma^{k}_{l,m}\cdot\sigma^{k}_{l,m} - a(\sigma^{k}_{l,m})^2 
+ a\sigma^{k+1}_{l+1,m}\sigma^{k-1}_{l-1,m}=A(u,y,l)(\sigma^{k}_{l,m})^2,\label{bl8:derivation2}
\end{equation}
where $A(u,y,l)$ is an arbitrary function. Since $\sigma^k_{l,m}=1$ (the case of $N=0$) satisfies
(\ref{bl5}) and (\ref{bl10}), it should satisfy (\ref{bl8:derivation2}) as well. Therefore we see
that $A$ must be $0$, which implies (\ref{bl8}).
\section*{Acknowledgements} 
This work is partially supported by JSPS Grant-in-Aid for Scientific Research No.~19340039,
21540067, 21656026, 22656027 and 23340037.


\end{document}